\documentclass[english,prb,twocolumn,superscriptaddress,showpacs,preprintnumbers,amsmath,amssymb]{revtex4}

\usepackage{graphicx}
\usepackage{dcolumn}
\usepackage{bm}
\usepackage{epsfig}
\newcommand{\ignore}[1]{}

\begin{document}

\title{Average Density of States in Disordered Graphene systems}

\author{Shangduan Wu}
\address{Hefei National Laboratory for Physical Sciences at
Microscale, University of Science and Technology of China, Hefei,
Anhui 230026, People's Republic of China}
\address{Department of Physics, University of Science and
Technology of China, Hefei, Anhui 230026, People's Republic of
China}

\author{Lei Jing}
\address{Hefei National Laboratory for Physical Sciences at
Microscale, University of Science and Technology of China, Hefei,
Anhui 230026, People's Republic of China}
\address{Department of Physics, University of Science and
Technology of China, Hefei, Anhui 230026, People's Republic of
China}

\author{Qunxiang Li}
\address{Hefei National Laboratory for Physical Sciences at
Microscale, University of Science and Technology of China, Hefei,
Anhui 230026, People's Republic of China}

\author{Q. W. Shi}
\email[Electronic address:]{phsqw@ustc.edu.cn}
\address{Hefei National Laboratory for Physical Sciences at
Microscale, University of Science and Technology of China, Hefei,
Anhui 230026, People's Republic of China}

\author{Jie Chen}
\address{Electrical and Computer Engineering, University of Alberta, AB T6G 2V4, Canada}%
\address{National Institute of Nanotechnology, Canada}

\author{Xiaoping Wang}
\address{Hefei National Laboratory for Physical Sciences at
Microscale, University of Science and Technology of China, Hefei,
Anhui 230026, People's Republic of China}

\author{Jinlong Yang}
\address{Hefei National Laboratory for Physical Sciences at
Microscale, University of Science and Technology of China, Hefei,
Anhui 230026, People's Republic of China}

\date{\today}

\begin{abstract}

In this paper, the average density of states (ADOS) with a binary
alloy disorder in disordered graphene systems are calculated based
on the recursion method. We observe an obvious resonant peak caused
by interactions with surrounding impurities and an anti-resonance
dip in ADOS curves near the Dirac point. We also find that the
resonance energy ($E_{r}$) and the dip position
($\varepsilon_{dip}$) are sensitive to the concentration of
disorders ($x$) and their on-site potentials ($v$). An linear
relation, $\varepsilon_{dip}$=$xv$, not only holds when the impurity
concentration is low but this relation can be further extended to
high impurity concentration regime with certain constraints. We also
calculate the ADOS with a finite density of vacancies and compare
our results with the previous theoretical results.

\end{abstract}


\maketitle

\section{\label{sec:level1}Introduction}

Graphene is a two-dimensional (2D) material with a single atomic
layer of graphite. The material has been fabricated firstly by
rubbing graphite layers against an oxidized silicon surface
recently.\cite{Science06-Nov} Due to the linear dispersion relation
of its electronic spectrum near the Dirac point, the electron
transport behavior of graphene at low energy range is essentially
determined by the massless relativistic Dirac's equations. Many
interesting properties of graphene have been studied experimentally
and theoretically by many research groups, including the unusual
quantum Hall effect,\cite{Nature05-Nov, Nature05-Zha,
Cond-mat05-Per, PRB06-Per} quantum minimal
conductivity,\cite{PRB06-Per, EPJ06-Kat, PRB07-Cse, PRL07-Zie,
PRB06-Gus, PRB07-Zie, PRL07-Ken}
ferromagnetism,\cite{AdvMater03-Han, PRB05-Voz} and
superconductivity.\cite{PRB97-Tam, JSoc97-Nak, JLowTemp00-Kop}

The disorder in graphenes can significantly impact their electronic
properties and has been studied extensively.\cite{J.Mol.Str.91-Ovc,
PRL92-Gon, PRB96-Cha, PRL00-Wak, PRB01-Wak, PRB01-Gon, JPCM01-Har,
JPSJ01-Mat, JPCS04-Har, PRL04-Dup, PRL04-Leh, PRB05-Voz} It is
believed that the interplay between disorders and electron-electron
interactions determine the low energy behavior of the electron in
the graphene system. Due to disorders, the average density states
(ADOS) increases at the Dirac points.\cite{PRB06-Per} The ADOS is
actually an important parameter to describe electronic structures,
especially in a disordered system. The minimum conductivity
($\sigma_{min}$), for instance, can be obtained by calculating the
diffusion constant ($D$) and ADOS ($\rho$) at the Fermi level
through the Einstein relation ($\sigma_{min}=\rho D$). To date,
various methods have been proposed to calculate ADOS and the local
density of states (LDOS) in various types of disordered graphenes,
such as Anderson disorder,\cite{PRB84-Hu} short-range potential
disorder,\cite{PRL07-Ken, PRB06-Skr, PRB07-Skr, PRB07-Weh,
Cond-mat06-Pog}, long-range potential disorder\cite{PRL07-Ken}, and
vacancy.\cite{PRB06-Per,PRL06-Pere,Cond-mat05-Per} However, these
calculation results are unreliable due to the limitations of
approximation, or these calculations are only suitable to address
electronic structure with a low impurity concentration. It is,
therefore, important to obtain the accurate electronic structures of
disordered graphenes.

Recently, the phenomenon of spectrum rearrangement has been studied
in a binary alloy disorder in disordered graphene. Due to the
limitation of the coherence potential approximation
(CPA),\cite{PRB07-Skr, PRB04-Skr} the corresponding results are
appropriate to address the electronic structures of graphene with an
extreme low impurity concentration. In this paper, we calculate the
ADOS with the similar system by using the recursion method. Our
numerical simulations provide the accurate ADOS for different
impurity concentrations. Moreover, our simulation can also be
generalized to study other types of disordered graphene. Though the
main features of ADOS with a binary alloy disorder in graphene can
be described using the resonant and anti-resonant states caused by
scattering of a impurity as reported in Ref. 20. Interestingly, we
observe that the impurity concentration ($x$) and the on-site
potential ($v$) have significant impact on the main features of
ADOS. The impurity concentration shifts the position of resonance
energy ($E_{r}$). An linear relation for the anti-resonance dip
shift ($\varepsilon_{dip}= xv$) with a relative low impurity
concentration can be extended to a high impurity concentration case
under certain conditions. Moreover, the ADOS with a finite
concentration of vacancies is also calculated using our proposed
approach and compared with the other previous theoretical results.

This paper is organized as follows. The tight-binding model of
graphene with a binary alloy disorder is given in Sec. II A. The
recursion method is introduced and its accuracy and applications are
explored in Sec. II B. The ADOS of graphene with different impurity
concentration\textbf{s} and on-site potentials are calculated and
the detailed discussions are given in Sec. III. Finally, we conclude
our contributions in Sec. IV.

\section{Computational Model and Method}

\subsection{\label{sec:level2}Model}

Fig.~1 shows the hexagonal lattice structure of a graphene and each
unit cell has two inequivalent atoms labeled A and B, respectively.
If we consider the contribution from the $\pi$ bond (one $\pi$
electron per atom) and the nearest interactions in the graphene, the
Hamiltonian based on the Wannier representation can be expressed as

\begin{eqnarray}
\hat{H}=\sum_{i}\varepsilon_{i}|i\rangle\langle
i|-t_{ij}\sum_{\langle i, j\rangle}|i\rangle\langle j|,
\end{eqnarray}

\noindent where, \emph{i} and \emph{j} denote the neighboring sites
on the lattice, $\varepsilon_{i}$ is the on-site energy, and
$t_{ij}$ is the nearest hopping energy ($t_{ij}$=$t$ and its value
is close to 2.7 eV in graphenes). For simplicity, here $t$ is scaled
to be 1.

\begin{figure}[htbp]
\begin{center}
\includegraphics[width=7.5cm]{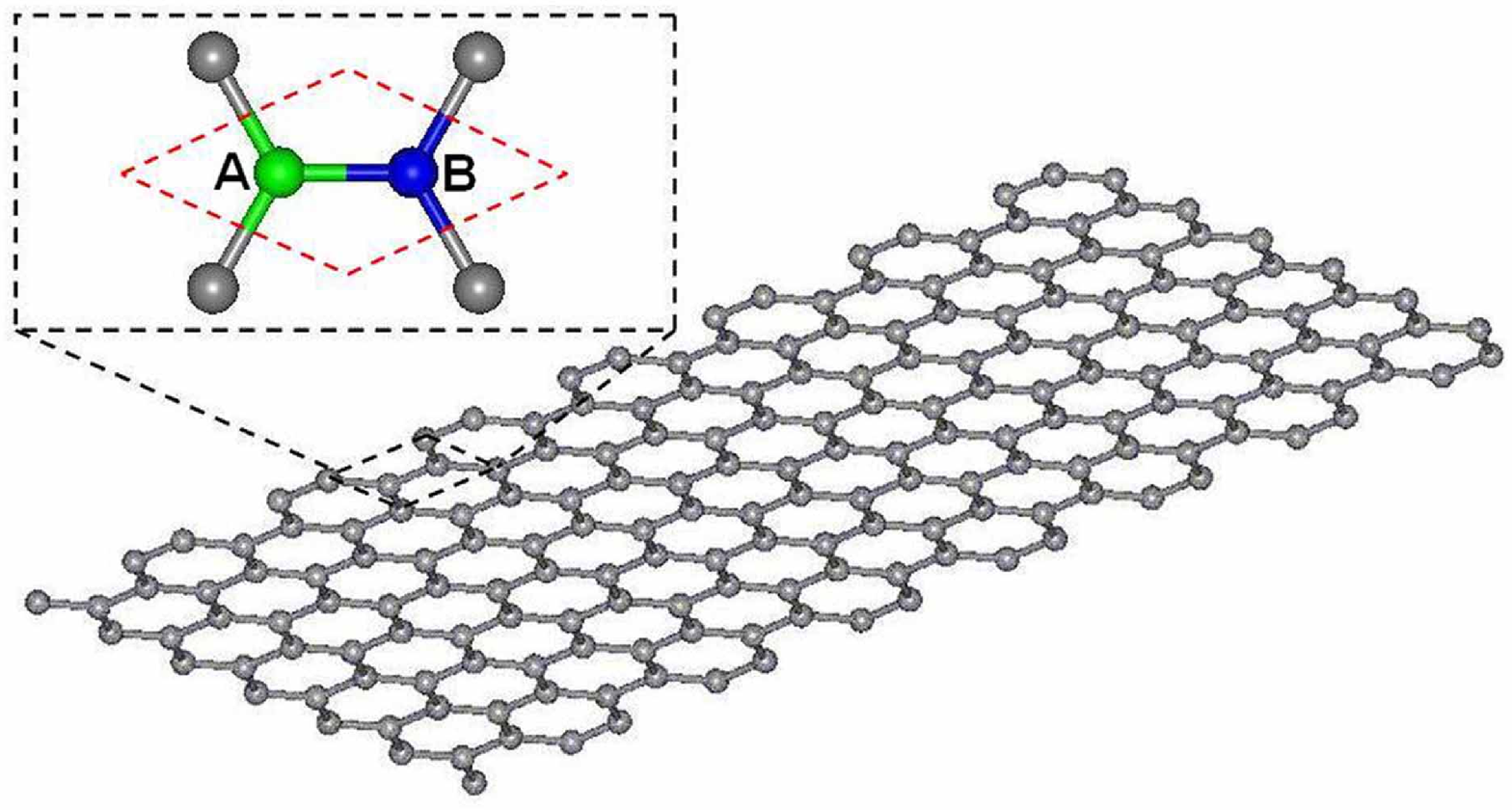}
\end{center}
\caption{(Color online) A honeycomb lattice of a graphene. Here, an
unit cell is outlined by the red dashed lines. Two inequivalent
atoms in the unit cell are labeled with A (the green sphere) and $B$
(the blue one), respectively.}
\end{figure}

With a given impurity concentration ($x$), the on-site energy
$\varepsilon_{i}$ for a binary alloy disordered system equals $v$
with probability $x$, or zero otherwise. This model, which is
attributed to Lifshiz,\cite{ Adv. Phys.64-Lif}, features the
absolutely random distribution of impurities in the space domain. In
our following calculations, we emphasize on studying the properties
of the ADOS in two cases. One is on-site potential ($v$) varies with
given disorder concentrations ($x$). In the other case, the
concentration ($x$) changes with given on-site potentials ($v$).

\subsection{The Recursion Method}

The Lanczos method\cite{JRNAB50-Lan} (so called the
Haydock-Heine-Kelly recursion method\cite{JPC72-Hay, JPC75-Hay}) is
a commonly used approaches to calculate the ADOS in disorder
systems. The essential idea of the recursion method is that the
Hamiltonian matrix is expressed using a tridiagonal representation
iteratively. After selecting a localized seed-state
($|f_{0}\rangle$), this recursion method generates a hierarchy of
states ($|f_{n}\rangle$) based on a defined orthogonal basis
recursively.

\begin{eqnarray}
|f_{n+1}\rangle=\hat{H}|f_{n+1}\rangle-\frac{\langle
f_{n}|\hat{H}|f_{n}\rangle}{\langle f_{n}|f_{n}
\rangle}|f_{n}\rangle-\frac{\langle f_{n}|f_{n}\rangle}{\langle
f_{n-1}|f_{n-1} \rangle}|f_{n-1}\rangle,
\end{eqnarray}

\noindent where, $n=0,1,2,\cdots$, and the recursive coefficients
are given by

\begin{eqnarray}
a_{n}=\frac{\langle f_{n}|\hat{H}|f_{n}\rangle}{\langle f_{n}|f_{n}
\rangle},b_{n}=\frac{\langle f_{n}|f_{n}\rangle}{\langle
f_{n-1}|f_{n-1} \rangle} (b_{0}=0, |f_{-1}\rangle=0).
\end{eqnarray}

In the orthogonal basis, the Hamilton matrix becomes

\begin{equation}
H=\begin{pmatrix}
a_{0} & b_{1} & 0 & 0 & \cdots \\
b_{1} & a_{1} & b_{2} & 0 & \cdots \\
0 & b_{2} & a_{2} & b_{3} & \cdots \\
0 & 0 & b_{3} & a_{3} & \cdots \\
\vdots & \vdots & \vdots & \vdots & \ddots.
\end{pmatrix}
\end{equation}

The diagonal elements in a Green's function matrix for a seed state
can be derived from Eq.~(4) based on the continuous-fraction method.

\begin{eqnarray}
G_{00}(E)&=&\langle f_{0}|\cfrac{1}{E-H}|f_{0}\rangle \nonumber \\
         &=&\cfrac{1}{E-a_{0}-\cfrac{b_{1}^{2}}{E-a_{1}-\cfrac{b_{2}^{2}}{E-a_{2}-\cfrac{b_{3}^{2}}{\ddots}}}}
\end{eqnarray}

\noindent and LDOS is defined as

\begin{eqnarray}
\rho_{local}(E)=\lim\limits_{\varepsilon\to
0^{+}}[-\frac{1}{\pi}G_{00}(E+i\varepsilon)].
\end{eqnarray}

We can then calculate ADOS easily using the following equation

\begin{eqnarray}
\rho_{aver}(E)=\frac{1}{M}\sum_{M}\rho_{local}(E),
\end{eqnarray}

\noindent where $M$ is the number of the samples. To terminate
continuous fractions, Eq.~(5) can be rewritten as

\begin{eqnarray}
G_{00}(E)=\cfrac{1}{E-a_{0}-\cfrac{b_{1}^{2}}{E-a_{1}-\cfrac{b_{2}^{2}}{\cfrac{\cdots}{E-a_{n}-t(E)}}}}.
\end{eqnarray}

In above equation, the terminating term can be written as

\begin{eqnarray}
t(E)=\cfrac{1}{2}\{(E-a_{\infty})-[(E-a_{\infty})^{2}-4b_{\infty}^{2}]\}.
\end{eqnarray}

The asymptotic value of the continuous-fraction coefficient pairs
$(a_{\infty},b_{\infty})$ in above equation can be obtained when $n$
gets large.

This recursion method has been adopted to investigate various
disorder systems.\cite{PRB94-Hay, PRB82-Sin, PRB89-Loh} In fact, an
infinite system can be approximated with periodic boundary condition
using this method and the numerical error of DOS is easy to
estimate.\cite{PRB94-Hay} If the system size ($L$) and its recursion
step ($N$) are large enough and corresponding positive broadening
width ($\varepsilon$) is reasonably small, the LDOS results should
be accurate. It is easy to get a reliable ADOS of a disorder system
by averaging over a large number of samples ($M$). For a perfect
graphene, two relations,  $\varepsilon\sim1/L$ and $\rho\sim1/L$,
hold approximately at the Dirac point as shown in Figs.~2 $(a)$ and
$(b)$, respectively. The results suggest that the DOS in a large
finite system is close to that in an infinite system. Of course, to
obtain a reliable ADOS result, a large recursion step $N$ is needed,
which should be larger than $L$, i.e. 2$L$ [Fig.~2$(c)$].

\begin{figure}[htbp]
\begin{center}
\includegraphics[width=7.5cm]{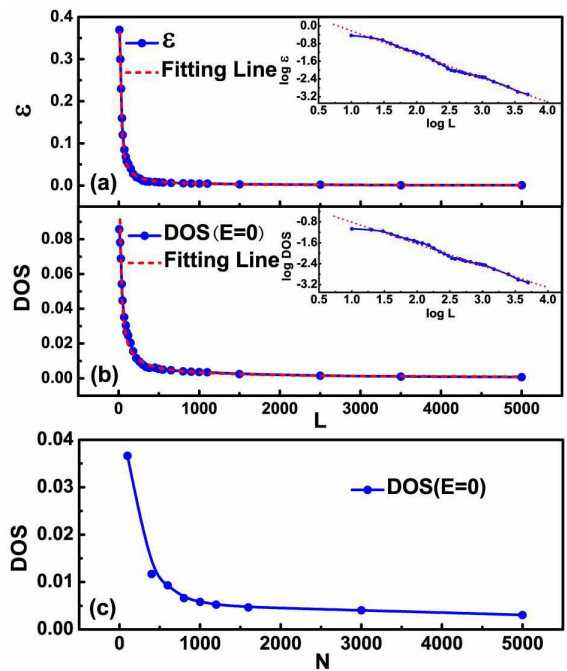}
\end{center}
\caption{(Color online) (a) $\varepsilon$ as a function of $L$ (the
blue solid line) and the corresponding fitting line shown with red
dashed line for a perfect graphene. (b) DOS ($\rho$) as a function
of $L$ (the blue solid line) and corresponding fitting line with the
red dashed line. (c) DOS ($\rho$) as a function of $N$ (the blue
solid line) at the Dirac point ($E=0$). Here, $L=800$.}
\end{figure}

The accuracy of the recursion method is examined by comparing the
calculated DOS of a perfect graphene with those obtained based on
the strict integral method in the first Brillouin zone. The
calculated DOS at several energy points are list in Table I.
Clearly, the results using the recursion method are reliable and
their relative errors are very small. In what follows, we calculate
the ADOS of a disorder graphene and the parameters, $L=800$, $N=2L$,
$\varepsilon=5\times10^{-3}$ and $M=1000$, are chosen.

\begin{table}
\caption{The calculated density of state in a perfect graphene using
both the recursion method and the strict integral method within the
first BZ. Here, $L=800$, $N=2L$ and $\varepsilon=5\times10^{-3}$.}
\begin{ruledtabular}
\begin{tabular}{cccccccc}
Energy              & $0.0$    & $0.4$    & $0.8$    & $1.2$    & $1.6$    & $2.0$    &\\
\hline
REM\footnotemark[1] & $0.0037$ & $0.0781$ & $0.1990$ & $0.2570$ & $0.1952$ & $0.1702$ & \\
INM\footnotemark[2] & $0.0037$ & $0.0781$ & $0.1998$ & $0.2529$ & $0.1938$ & $0.1695$ & \\
Error               & $<0.0001$& $<0.0001$& $<0.002$ & $<0.002$ & $<0.002$ & $<0.002$   & \\
\end{tabular}
\end{ruledtabular}
\footnotetext[1]{The recursion method} \footnotetext[2]{The strict
integral method}
\end{table}

\section{Results and Discussion}

Fig.~3(a) shows several calculated ADOS of graphene with the fixed
impurity concentration $x=10\%$, but different on-site potentials
($v$). Our results clearly show that ADOS remains the similar shape
in a high energy region. However, the ADOS near the Dirac point
changes remarkably for different on-site potential ($v$). Two
obvious features are observed. (1) An obvious resonance peak appears
in the ADOS curves when $v\geq3$. Its energy position or the
resonance energy ($E_{r}$) shifts towards the Dirac point when the
on-site potential ($v$) increases. This peak is attributed to a
resonance state. A very similar feature was also observed in the
LDOS when there is a single impurity or a low impurity
concentration.\cite{PRB06-Skr, PRB07-Skr} (2) There is a dip near
the Dirac point, which results from an antiresonance state. The
position of this anti-resonance dip ($\varepsilon_{dip}$) shifts due
to the presence of impurities. As shown in Fig. 3(b), the linear
relation, $\varepsilon_{dip}$=$xv$, holds when the on-site potential
is relatively small, but the dip disappears when $v$ becomes larger
$ (i.e. v>3)$. These results show that the relation
($\varepsilon_{dip}$=$xv$) is correct when $v\ll v_{dip}$, while
this relation does not hold for a system with on-site potential
$v>v_{dip}$.

To quantitatively explore the impurity concentration how to shift
the position of resonance energy ($E_r$), we compare the result for
the finite concentration with that for a single impurity. The
calculated results are shown in Fig.~3(c). Here, we adopted the
recursion method and the effective-mass approximation method
(EMA)\cite{PRB06-Wang} to determined the position of resonance
energy ($E_r$) of LDOS at impurity site for a single impurity in the
graphene system respectively. We also calculate resonance energy
($E_r$) of ADOS for the finite impurity concentration by using the
recursion method.

The Green's function in a perfect graphene based on the
effective-mass approximation is expressed as

\begin{eqnarray}
G_{0}(E)&=&\lim\limits_{\varepsilon\to
0^{+}}\cfrac{S}{\pi}\int_{0}^{k_{c}}\cfrac{(E+i\varepsilon)k}{(E+i\varepsilon)^{2}+(3tak/2)^{2}}dk
\nonumber \\
        &=&\cfrac{\sqrt{3}}{3\pi
        t^{2}}E\ln|\cfrac{E^{2}}{9t^{2}a^{2}k_{c}^{2}/4-E^{2}}|-i\cfrac{\sqrt{3}}{3t^{2}}|E| \nonumber \\
        &=&\cfrac{\sqrt{3}}{3\pi}E\ln|\cfrac{E^{2}}{9a^{2}k_{c}^{2}/4-E^{2}}|-i\cfrac{\sqrt{3}}{3}|E|,
\end{eqnarray}

\noindent where $S=3\sqrt{3}a^{2}/2$ is the area of a unit cell in
real space. $a$ is lattice constant. $k_{c}$ is cutoff wave vector
and is set to be $2.13/a$.\cite{PRB06-Wang} For a simple case, the
Green's function of a single impurity graphene is expressed as

\begin{eqnarray}
G(E)&=&\cfrac{G_{0}(E)}{1-vG_{0}(E)}=(\cfrac{\sqrt{3}}{3\pi}E\ln|\cfrac{E^{2}}{9a^{2}k_{c}^{2}/4-E^{2}}| \nonumber \\
    &&-\cfrac{v}{3\pi}E^{2}\ln^{2}|\cfrac{E^{2}}{9a^{2}k_{c}^{2}4-E^{2}}|-\cfrac{v^{2}}{3}E^{2}-i\cfrac{\sqrt{3}}{3}|E|) \nonumber \\
    &&/([1-\cfrac{\sqrt{3}}{3\pi}E\ln|\cfrac{E^{2}}{9a^{2}k_{c}^{2}/4-E^{2}}|]^2+\cfrac{v^{2}}{3}E^{2}).
\end{eqnarray}

LDOS at the impurity site is easy to obtain by
\begin{eqnarray}
\rho_{imp}(E)=\cfrac{\sqrt{3}}{3\pi}\cfrac{|E|}{[1-\cfrac{\sqrt{3}}{3\pi}E\ln|\cfrac{E^{2}}{9a^{2}k_{c}^{2}/4-E^{2}}|]^2+\cfrac{v^{2}}{3}E^{2}}.
\end{eqnarray}

The resonant energy ($E_{r}$) in this case can be defined using the
Lifshits equation as

\begin{eqnarray}
1=vReG_{0}(E_{r})=\cfrac{\sqrt{3}v}{3\pi}E_{r}\ln|\cfrac{E^{2}_{r}}{9a^{2}k_{c}^{2}/4-E^{2}_{r}}|.
\end{eqnarray}

\begin{figure}[htbp]
\begin{center}
\includegraphics[width=7.5cm]{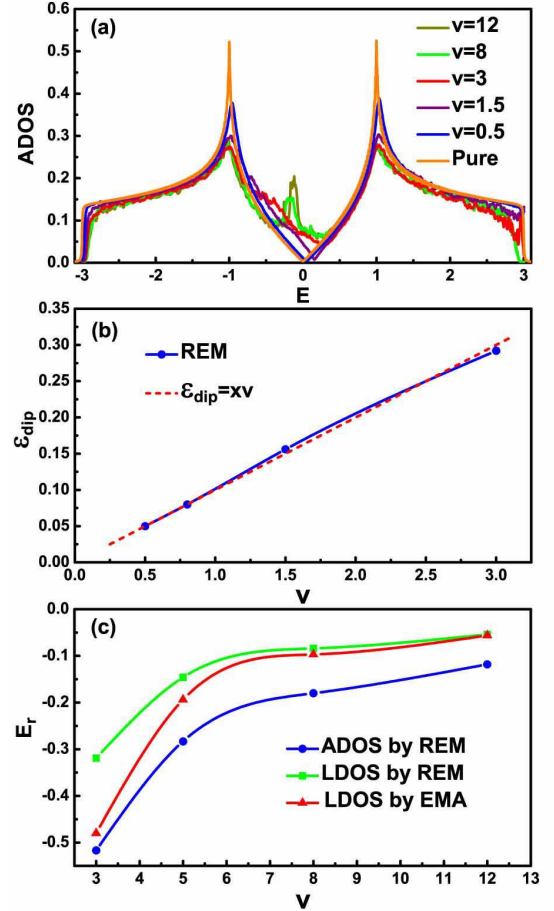}
\end{center}
\caption{(Color online) (a) ADOS ($\rho$) of a graphene with
impurity concentration $x=10\%$ and various on-site potentials
($v$). (b) The energy shift of the anit-resonance state
($\varepsilon_{dip}$) as a function of the on-site potential ($v$)
(the blue solid line). The red dashed line stands for the linear
relation, $\varepsilon_{dip}=xv$. (c) The resonance energy ($E_{r}$)
as a function of the on-site potential ($v$). The blue, green, and
red lines stand for the ADOS for a graphene with impurity
concentration $x=10\%$. ADOS is calculated using the recursion
method, LDOS of a single impurity is computed using the recursion
method, and LDOS of a single impurity is obtained using EMA,
respectively.}
\end{figure}

When on-site potential ($v$) increases as shown in Fig.~3(c), the
position ($E_r$) of resonance state shifts toward to the Dirac point
as shown in Fig.~3(a). $E_{r}$ for the finite concentration is quite
different from that for a single impurity. The difference is caused
by the multiply scattering among impurities. Our simulation implies
that the multiple scattering process is important for finite
concentration and its contribution neglected in some approximation
methods is not available. Note that the positions of $E_{r}$ for a
single impurity determined by the recursion method and by the
effective approximation respectively are still quite difference at a
small on-site potential and coincide each other at a large on-site
potential. It is because the Green's function obtained by Eq.~(10)
is validate only for low-energy (very close to the Dirac point)
region.

\begin{figure}[htbp]
\begin{center}
\includegraphics[width=7.5cm]{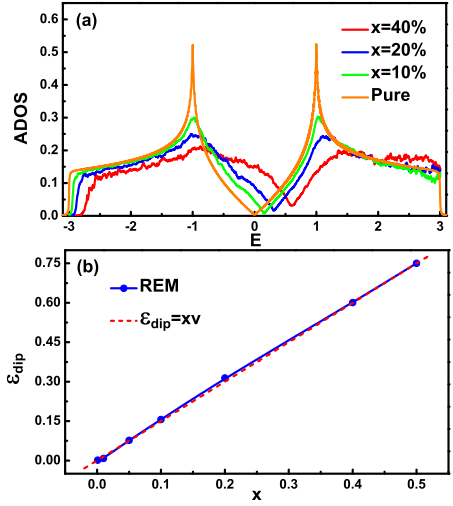}
\end{center}
\caption{(Color online) (a) ADOS ($\rho$) of a graphene with various
(number of disorders)impurity concentrations ($x$). Here, on-site
potential is set to be $1.5$ ($v=1.5$).  (b) The energy shift
($\varepsilon_{dip}$) as a function of impurity concentrations ($x$)
(the blue solid line), for on-site potential $v=1.5$. The red dashed
line shows the linear relation: $\varepsilon_{dip}=xv$.}
\end{figure}

Fig.~4(a) presents the calculated ADOS of a graphene with different
impurity concentrations($x$) while the on-site potential ($v$) is
set to be $1.5$.  When the on-site potential ($v$) is small, the
anti-resonance dip exists, but there is no obvious resonance peak.
Interestingly, we find that the linear relation,
$\varepsilon_{dip}$=$xv$, still holds even for a large impurity
concentration, though the multiply scattering among impurities is
strong. Generally, this linear relation can be easily understood by
using virtual crystal approximation(VCA) which simply predicts a
energy level at which two bands coincide is shifted by
$\varepsilon_{dip}$=$xv$. The prediction of VCA is reasonable for a
low impurity concentration and a small on-site potential, but it
fails for large values. It is also found that the van Hove
singularity disappears when the impurity concentration is about
$40\%$. This reflects that the phenomenon of the spectrum
rearrangement happens in the large impurity concentration.

In order to get better understanding of the linear relation,
$\varepsilon_{dip}$=$xv$, here, we present a qualitative discussion
based on the coherent potential approximation (CPA).The Self-energy
$\Sigma$ can be obtained by neglecting multiple scattering process
among the impurities as

\begin{eqnarray}
\Sigma=\cfrac{xv}{1-vG_{0}(E-\Sigma)}
\end{eqnarray}

The energy shift $\varepsilon_{dip}$ can be determined by
substituting $E-\Sigma$ =$i\kappa$ ($\kappa>0$ is real). Thus we
have,

\begin{eqnarray}
\varepsilon_{dip}\equiv Re\Sigma=\cfrac{xv(1-\cfrac{\sqrt{3}\kappa
v}{3})}{(1-\cfrac{\sqrt{3}\kappa
v}{3})^{2}+\cfrac{1}{3\pi_{2}}\kappa^{2}v^{2}\ln^{2}|\cfrac{\kappa^{2}}{9a^{2}k_{c}^{2}/4-\kappa^{2}}|}.
\end{eqnarray}

\noindent $\kappa$ is calculated by the following equation,

\begin{eqnarray}
Im\Sigma=-\kappa=\cfrac{\cfrac{\sqrt{3}}{3\pi}\kappa
v^{2}\ln|\cfrac{\kappa^{2}}{9a^{2}k_{c}^{2}/4-\kappa^{2}}|}{(1-\cfrac{\sqrt{3}\kappa
v}{3})^{2}+\cfrac{1}{3\pi_{2}}\kappa^{2}v^{2}\ln^{2}|\cfrac{\kappa^{2}}{9a^{2}k_{c}^{2}/4-\kappa^{2}}|}.
\end{eqnarray}

\begin{figure}[htbp]
\begin{center}
\includegraphics[width=7.5cm]{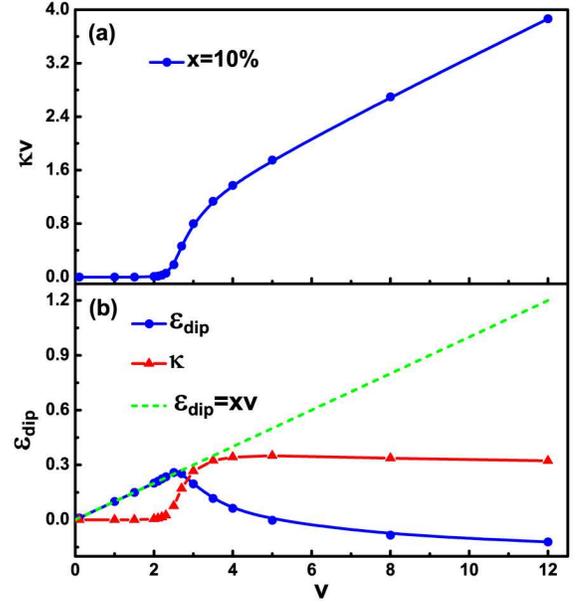}
\end{center}
\caption{(Color online) (a) Parameter $\kappa v$ as a function of
on-site potential ($v$), where impurity concentrations is $x=10\%$.
(b) The energy shift ($\varepsilon_{dip}$) as the function of
on-site potential ($v$). The blue and red solid lines stand for the
results calculated using the recursion method and CPA, respectively.
In the calculations, we choose $x=10\%$. The green dashed line
indicates the linear relation, $\varepsilon_{dip}=xv$.}
\end{figure}

If $\kappa v\ll 1$, Eq.~(15) can be simplified to
$\varepsilon_{dip}$=$xv$. For a system with a given impurity
concentration e.g.$10\%$,  Fig.~6(a) clearly shows that $\kappa v$
is a small value for $v$ less than 2.5. If $\kappa v\ll 1$, the
relation, $\varepsilon_{dip}$=$xv$, is correct as shown in
Fig.~3(b). However, the calculated curve of $\varepsilon_{dip}$
using CPA does not match the linear relation
($\varepsilon_{dip}$=$xv$) when $v$ is larger than 2.5.  The
possible reason is that multiple scattering among the impurity
cannot be neglected in the CPA calculations when $v>2.5$. With given
the impurity concentration $10\%$, based on the condition of
$|Re\Sigma|=|Im\Sigma|$, CPA calculation estimate\textbf{s} roughly
the threshold value $v_{dip}\approx 2.9$ of on-site potential
($v_{dip}$) to guarantee the linear relation. This result is
somewhat underestimated by looking at Fig.~3(a), for instance, the
value of $v_{dip}$ should be larger than 3. For the system with a
given on-site potential ($v$), Fig.~6(a) shows that the results
obtained using the CPA method are different from those obtained
using the recursion method. In the case of $v=1.5$, the inequality
$\kappa v\ll 1$ is satisfied only for $x\leq 0.2$. With small
on-site potential, we find that the relation
$\varepsilon_{dip}$=$xv$ still holds even when impurity
concentration is close to $50\%$. This observation indicates that
the CPA is not applicable to obtain the reliable results of graphene
when impurity number is large because which neglects the scattering
among impurities.

\begin{figure}[htbp]
\begin{center}
\includegraphics[width=7.5cm]{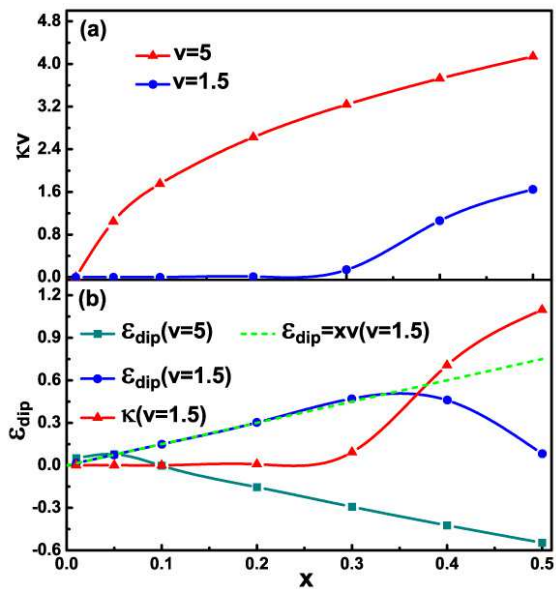}
\end{center}
\caption{(Color online)(a) Parameter $\kappa v$ as a function of
impurity concentration ($x$) when $v=1.5$ or $v=5$. (b) The energy
shift ($\varepsilon_{dip}$) as a function of impurity concentration
($x$). The blue and dark cyan solid lines stand for the case when
$v=1.5$ and $v=5$, respectively. The red solid line indicates
$\kappa$ as a function of impurity concentration ($x$) for $v=1.5$
case. The green dashed line shows the linear relation of
$\varepsilon_{dip}=xv$.}
\end{figure}

The threshold impurity concentration ($x_{dip}$) for $v=1.5$ is
about $37\%$ when the liner relationship is correct, as show in
Fig.~6(b). When the on-site potential ($v$) becomes large enough,
$x_{dip}$ is extremely small. This result is consistent with our
calculated results shown in Fig.~5. There is a visible dip only for
extreme small impurity concentrations. Clearly, $x_{dip}$ is
sensitive to the on-site potentials. When the inequality $\kappa
v\ll 1$ does not hold anymore, $x_{dip}$ cannot be obtained through
the simple CPA calculations.\cite{PRB07-Skr} Previous theoretical
studies \cite{ PRB07-Skr,PRB04-Skr} have predicted that the CPA with
single-site scattering may fail to predict the dip. Our results
obtained by the recursion method prove that the CPA is valid only if
($\kappa v\ll 1$). The linear relation, $\varepsilon_{dip}$=$xv$,
can be further extended to system with high impurity concentration
and small on-site potential though the CPA fails to work.

\begin{figure}[htbp]
\begin{center}
\includegraphics[width=7.5cm]{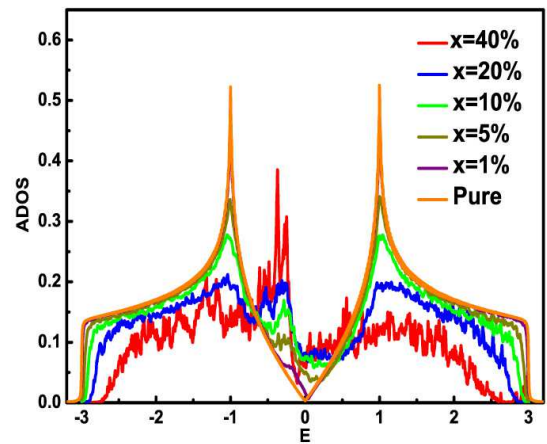}
\end{center}
\caption{(Color online) The calculated ADOS ($\rho$) with various
impurity concentration ($x$) in a graphene. Here, $v=5$.}
\end{figure}

For large on-site potentials as shown in Fig.~7, there are obvious
resonance peaks. The height of the peak increases significantly as
the number of disorders increases. However, the position of
resonance peak does not shift obviously at high impurity
concentration. These results suggest that $E_{r}$ is determined by
the on-site potential ($v$) and impurity concentration ($x$).
However, when the on-site potential is large, $E_r$ does not depend
obviously on the impurity concentration even $x$ has large value. It
is also found that the van Hove singularity disappears when the
impurity concentration is about $20\%$.

The disordered graphene with a finite density of vacancies can be
modeled by setting the on-site potential $v$ to a very large
value.\cite{PRB07-Weh, PRB06-Wang} Fig.~8 shows the calculated ADOS
of graphene systems with vacancies ($v=1000$) based on the recursion
method. There is a clear sharp peak near the Dirac point, which is
in line with the numerical results obtained using the stochastic
recusive method.\cite{PRL06-Pere}This sharp peak can be fit very
well by the Lorentz distribution. However, the precious theoretical
calculation, based on the full Born approximation
(FBA),\cite{PRB06-Per} predicted the value of ADOS should be exactly
zero. Its error may occur because the multiple scattering is ignored
in the FBA calculation. Meanwhile the calculation based on the
CPA\cite{Cond-mat05-Per} and the full self-consistent Born
approximation (FSBA)\cite{PRB06-Per} also fails to produce
observable resonance peaks. These results indicate that there are
some limitations in the CPA, FBA and FSBA methods near the Dirac
point. The correctness or accuracy of these methods need to be
further examined.

\begin{figure}[htbp]
\begin{center}
\includegraphics[width=7.5cm]{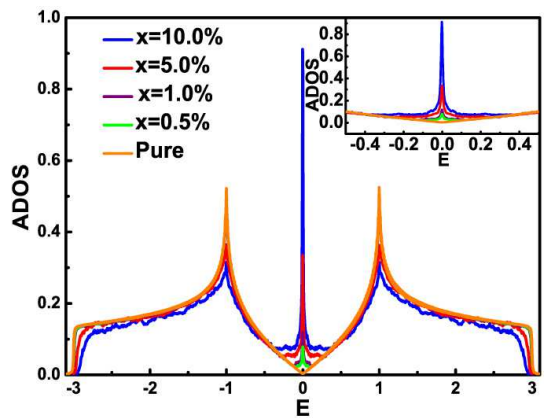}
\end{center}
\caption{(Color online) The calculated ADOS ($\rho$) of graphene
with different concentrations of vacancies. Here, we chose
$v=1000$.}
\end{figure}

\section{Conclusion}

In this paper, these ADOS of  a binary alloy disorder in disordered
graphene are calculated using the recursion method. The
applicability and accuracy of the recursion method are addressed. It
is found that the shape of the resonance peak and the position of
the anti-resonance dip are sensitive to impurity concentration ($x$)
and on-site potential ($v$). The linear relation,
$\varepsilon_{dip}$=$xv$, can be derived based on the CPA when
$\kappa v\ll 1$. This relation is able to explain the position shift
of anti-resonance dips at low or high impurity concentration for
small on-site potential case. For large $v$, either the CPA or
Eq.~(14) fails except for extremely low impurity concentration
($\kappa v\ll 1$). The main reason for the failure of the CPA is
that the scattering among impurities is neglected. By setting $v$ to
be a huge value, the model can be used to simulates finite
concentration of vacancies in a graphene. The resonance peak of ADOS
at the dirac point is found.

\section*{Acknowledgments}
This work is partially supported by the National Natural Science
Foundation of China (Grant nos. 10574119, 20773112, 10674121 and
50121202). The research is also supported by National Key Basic
Research Program under Grant No. 2006CB922000, Jie Chen would like
to acknowledge the funding support from the Discovery program of
Natural Sciences and Engineering Research Council of Canada under
Grant No. 245680.

\end{document}